\documentclass{ws-ijmpc}
\usepackage{amsmath}
\begin{document}
\bibliographystyle{elsart}

\markboth{T. P\"oschel, N.V. Brilliantov, and A. Formella}
{Granular gas cooling and relaxation to the steady state}

\catchline{}{}{}{}{}

\title{GRANULAR GAS COOLING AND RELAXATION TO THE STEADY STATE IN REGARD TO THE OVERPOPULATED TAIL OF THE VELOCITY DISTRIBUTION}

\author{THORSTEN P\"OSCHEL}

\address{Charit\'e, Augustenburger Platz, 10439 Berlin, Germany\\
thorsten.poeschel@charite.de}

\author{NIKOLAI V. BRILLIANTOV }

\address{Institute of Physics, University of Potsdam, Am Neuen Palais
  10, 14469 Potsdam, Germany \\
nbrillia@agnld.uni-potsdam.de}

\author{ARNO FORMELLA}

\address{Universidad de Vigo, Department of Computer Science, Edificio Polit\'ecnico,\\ 32004 Ourense, Spain\\
formella@ei.uvigo.es}

\maketitle

\begin{history}
\received{\today}
\revised{Day Month Year}
\end{history}

\begin{abstract}
We present a universal description of the velocity distribution
function of granular gases, $f(v)$, valid for both, small and
intermediate velocities where $v$ is close to the thermal velocity and
also for large $v$ where the distribution function reveals an
exponentially decaying tail. By means of large-scale Monte Carlo
simulations and by kinetic theory we show that the deviation from the
Maxwell distribution in the high-energy  tail leads to small but detectable variation of the 
cooling coefficient and  to extraordinary large relaxation time.

\end{abstract}

\ccode{PACS Nos.: 47.70.-n, 51.10.+y}

\section{Introduction}
Granular gases are characterized by a certain loss of kinetic energy
according to the dissipative properties of particle collisions. The
particle velocities before a collision, $\vec{v}_{1/2}$, and after,
$\vec{v}_{1/2}^{\,\prime}$, are related by the collision rule
\begin{equation}
  \label{eq:eps}
  \begin{aligned}
    \vec{v}_{1}^{\,\prime} & = \vec{v}_{1} -  \frac{1+\varepsilon}{2} \left(\vec{v}_{12} \cdot \vec{e} \, \right)\vec{e}\\
    \vec{v}_{2}^{\,\prime} & = \vec{v}_{2} +  \frac{1+\varepsilon}{2} \left(\vec{v}_{12} \cdot \vec{e} \, \right)\vec{e}
  \end{aligned}
\end{equation}
with $\vec{v}_{12}\equiv\vec{v}_1-\vec{v}_2$ and the unit vector  $\vec{e}
\equiv\left(\vec{r}_1-\vec{r}_2\right)/\left|\vec{r}_1-\vec{r}_2\right|$ at
the moment of the collision. 

In the absence of external excitation, a homogeneously initialized
granular gas stays homogeneous during the first part of its evolution,
called {\em homogeneous cooling state (HCS)}. In later stages of its
evolution, hydrodynamics instabilities develop leading to vortex
formation, that is, spacial correlations of the vectorial velocity
field,\cite{BritoErnst:1998} and finally cluster formation due to a
pressure instability.\cite{GoldhirschZanetti:1993} In this paper we
restrict ourselves to the homogeneous cooling state. It should be
mentioned that heated granular gases with a Gaussian thermostat are
equivalent to the HCS\cite{MontaneroSantosTailsGG:1999}, therefore,
the presented results apply also for such systems.

According to the steady loss of energy, the velocity distribution
function, $f\left(\vec{v},\tau\right)$, depends on time $\tau$
(measured in units of collisions per particle). Its shape can be
described by the time-independent function $\tilde{f}(\vec{c}\,)$
via\cite{EsipovPoeschel:1995}
\begin{equation}
  \label{eq:defScalf}
  f\left(\vec{v},\tau\right)=\frac{n}{v_T^3(\tau)} \tilde{f}\left(\vec{c}\,\right)
\end{equation}
where
\begin{equation}
  \label{eq:defScalf1}
  \vec{c} \equiv \frac{\vec{v}}{v_T(\tau)}\qquad\mbox{and}\qquad v_T(\tau)\equiv\sqrt{2T(\tau)}\,.
\end{equation}
The evolution of the granular temperature $T(\tau)$ is given by Haff's law\cite{Haff:83} (with particle mass $m_i=1$),
\begin{equation}
\label{eq:dTdt}
\frac{dT}{d\tau}= -2\gamma T\,,~~~~\mbox{i.e.,}\,~~~~~T(\tau)=T(0)\exp(-2\gamma\tau)\,.
\end{equation}
Here $\gamma$ is the cooling coefficient, addressed in detail in the
next section. 

\section{Universal velocity distribution function}

For small and intermediate velocities, $\left|\vec{v}\right|\approx
v_T$, that is, $\left|\vec{c}\,\right|\approx 1$, the scaled velocity
distribution function $\tilde{f}(\vec{c}\,)$ is close to a Gaussian
with small deviations that can be characterized by the first
non-trivial coefficient, $a_2$, of a Sonine polynomials expansion
around the leading Gauss
distribution,\cite{GoldshteinShapiro1:1995,NoijeErnst:1998}
\begin{equation}
  \label{eq:Sonexp}
  \tilde{f}(c)=  \pi^{-3/2}\exp\left(-c^2\right) \left[ 1 + a_2 S_2\left(c^2\right)\right]~~~~\mbox{with}~~~~S_2\left(c^2\right)=\frac{1}{1}c^4 - \frac{5}{2}c^2 +\frac{15}{8}\,
\end{equation}
and 
\begin{equation}
a_2(\varepsilon)= \frac{16(1-\varepsilon)\left(1-2\varepsilon^2\right)}{81-17\varepsilon+30\varepsilon+30^2(1-\varepsilon)}\,.
\label{eq:a2NE}
\end{equation}

For large kinetic energy, $c \gg 1$, the distribution function decays
exponentially slow, that is, the distribution function is
overpopulated with respect to the Gauss distribution one would expect
for a molecular gas,\cite{EsipovPoeschel:1995,NoijeErnst:1998}
\begin{equation}
\label{eq:tail}
\tilde{f}(c) =  B e^{-bc} \,~~~~~\mbox{with}~~~~~~
b=\frac{3\pi}{\mu_2}
\end{equation}
with the second moment of the dimensionless collision  integral
\begin{equation}
\label{eq:mu2}
\mu_2\equiv -\int d\vec{c}_1 c_1^2 \tilde{I}\left(\tilde{f},\tilde{f}\right) =\sqrt{2\pi}\left(1-\varepsilon^2\right)\left[1+\frac{3}{16}a_2\right]
\end{equation}
and with 
\begin{equation}
\label{eq:I_scale}
\tilde{I}\left(\tilde{f},\tilde{f}\right) = \int d\vec{c}_2\int d\vec{e}\,\, \Theta\!\left(-\vec{c}_{12}\cdot\vec{e}\,\right) \left|\vec{c}_{12}\cdot\vec{e}\,\right| \left[\frac{1}{\varepsilon^2}\tilde{f}\left(\vec{c}_1^{\,\prime\prime}\right) \tilde{f}\left(\vec{c}_2^{\,\prime\prime}\right) - \tilde{f}\left(\vec{c}_1\right) \tilde{f}\left(\vec{c}_2\right)\right]
\end{equation}
where $\vec{c}_{1/2}^{\,\prime\prime}$ are the pre-collisional scaled velocities.

There is a transition region between the ranges of validity of the
Sonine expansion ($c\approx 1$) and the expression for the tail ($c\gg
1$) where the distribution function was quantified in terms of
complicated implicit expressions.\cite{GoldhirschetalNLP:2003} For
practical computations, however, a much simpler approach may be
sufficient: It was shown recently that the simple combination of the
shown Sonine expansion and the exponential tail,
\begin{equation}
\label{eq:ftentat}
\tilde{f}(c)=\left\{
\begin{aligned}
  A c^2 e^{-c^2} \left[ 1 + a_2 S_2(c^2) \right] & \mbox{~~~~for} & c<c_*\\
  B c^2 e^{-bc}  & \mbox{~~~~for} & c\ge c_*
\end{aligned}
\right.
\end{equation}
with the parameters $A$, $B$, and the transition velocity $c_*$ agrees
with large-scale DSMC simulation up to a very good accuracy in the
entire range of velocities.\cite{PoeschelBrilliantovFormella:2006} The
parameters $A$, $B$, and $c_*$ follow from the conditions of
smoothness
\begin{equation}
  \label{eq:der_and_func}
  \begin{aligned}
    \tilde{f}(c_*+0) &= \tilde{f}(c_*-0)\\
    \left.\frac{d\tilde{f}}{dc}\right|_{c_*+0} &= \left.\frac{d\tilde{f}}{dc}\right|_{c_*-0}
  \end{aligned}
\end{equation}
and normalization:
\begin{eqnarray}
\label{eq:eq_for_cstar}
c_*&=&\frac{b}{2} +\frac{a_2 \left( 2 c_*^3 -5 c_* \right)}{2 \left[1+a_2 S_2\left(c_*^2\right)\right]}\\
\label{eq:def_A}
\frac{1}{A} &=& \frac{k(c_*)}{b^3} \! \left[ 2 + b c_* \left(2\! +\! b c_* \right) \right] e^{-bc_*}
\!+\! \frac{\sqrt{ \pi} }{4} {\rm     Erf} (c_*)  
\!-\!\frac{c_*}{8}\!  \left[4\!+\!a_2 c_*^2\!\left(2 c_*^2\! -\!5\right)\right]\! e^{-c_*^2}   \\[0.2cm]
\label{eq:def_B}
B &=& A k(c_*)
\end{eqnarray}
with
\begin{equation}
\label{eq:def_k}
k(c_*) \equiv  e^{-c_*^2 + b c_*} \left[ 1 + a_2 S_2 \left( c_*^2 \right) \right] \, .
\end{equation}
Solving the fifth order equation (\ref{eq:eq_for_cstar}) numerically
for $c_*$, we obtain $A$, $k$ and finally $B$.\cite{PoeschelBrilliantovFormella:2006} 

\section{Impact of the high-velocity tail on the coefficient of cooling}
By means of the distribution function, Eq. (\ref{eq:ftentat}), and its
parameters $A$, $B$, $c_*$ depending on the coefficient of restitution
$\varepsilon$, we can quantify the impact of the exponential tail on
the cooling coefficient $\gamma$, which is one of the most important
characteristics of a granular gas. Using the standard
analysis,\cite{BrilliantovPoeschelOUP} one can calculate the cooling
coefficient for the homogeneous cooling state, when the stationary
velocity distribution is achieved:
\begin{equation}
\label{eq:gam_gen}
\gamma=\left(1-\varepsilon^2\right)\frac{J_2}{12J_0} \, ,
\end{equation}
where the coefficients $J_k$ depend on the velocity distribution function,  
\begin{equation}
\label{eq:kin_int}
J_k =  \int  d \vec{c}_1   d \vec{c}_2   \int  d \vec{e} \,
 \Theta(-\vec{c}_{12} \cdot \vec{e} \, ) |\vec{c}_{12} \cdot \vec{e}\, |^{k+1} 
 \tilde{f}(c_1) \tilde{f}(c_2)  \, .
\end{equation}
If we neglect the exponential tail, we obtain the energy decay rate\cite{NoijeErnst:1998}
\begin{equation}
  \label{eq:gamma_0}
  \gamma_0(\varepsilon)=\frac{1-\varepsilon^2}{6}\frac{1+\frac{3}{16}
  \,a_2(\varepsilon)}{1-\frac{1}{16} \,a_2(\varepsilon)} \, . 
\end{equation}

Complementary, we determine the cooling coefficient numerically, using
{\em Direct Simulation Monte Carlo}
(DSMC)\cite{Bird:1994,MontaneroSantos:1996}. We initialized a granular
gas of $N=10^8$ particles with velocities drawn from a Gauss
distribution at $T(0)=1$ and simulated until the particle velocities
reached the limit of the double precision number representation, i.e.,
until $T\approx 10^{-23}$. For $\varepsilon=0.9$ this corresponds to a
total of $5\times 10^{10}$ collisions or 1,000 collisions per
particle.


For different $\varepsilon$ we recorded the temperature $T_{\rm
DSMC}(\tau)$. Then $\gamma_{\rm DSMC}$ was determined by fitting
$T_{\rm DSMC}(\tau)$ for $\tau\gg 1$ to its asymptotic law, $T_{\rm
DSMC}\propto\exp(-2\,\gamma_{\rm DSMC}\,\tau)$, Eq. (\ref{eq:dTdt}).

\begin{figure}[b]
\centerline{\includegraphics[width=8cm,clip]{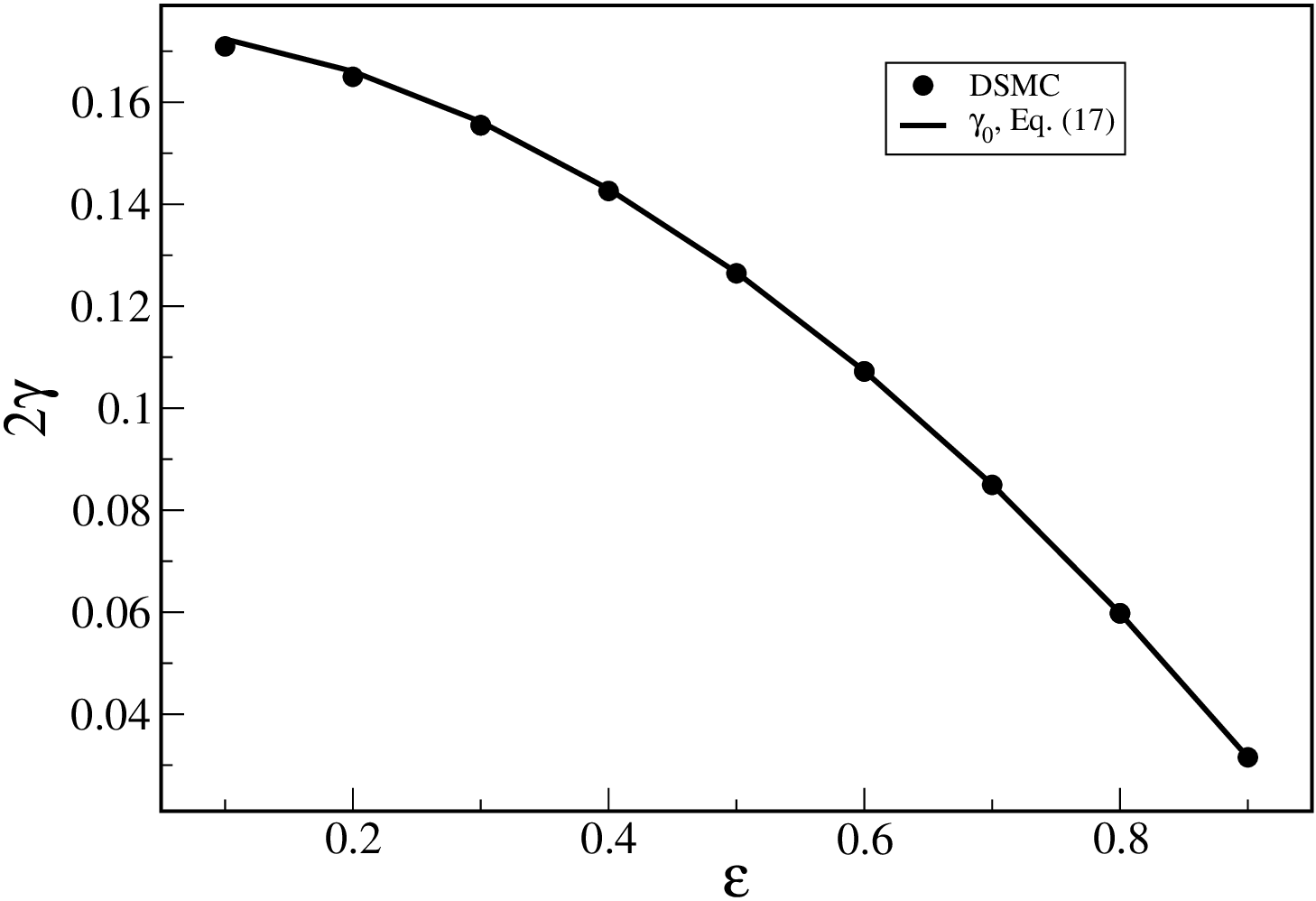}}
\caption{The cooling coefficient $\gamma(\varepsilon)$ as a function
  of the coefficient of restitution as obtained by Direct Simulation
  Monte Carlo (DSMC) (points) and due to Eq. (\ref{eq:gamma_0}),
  $\gamma_0$. }
\label{fig:gamma0}
\end{figure}

\begin{figure}[t]
\centerline{\includegraphics[width=8cm,clip]{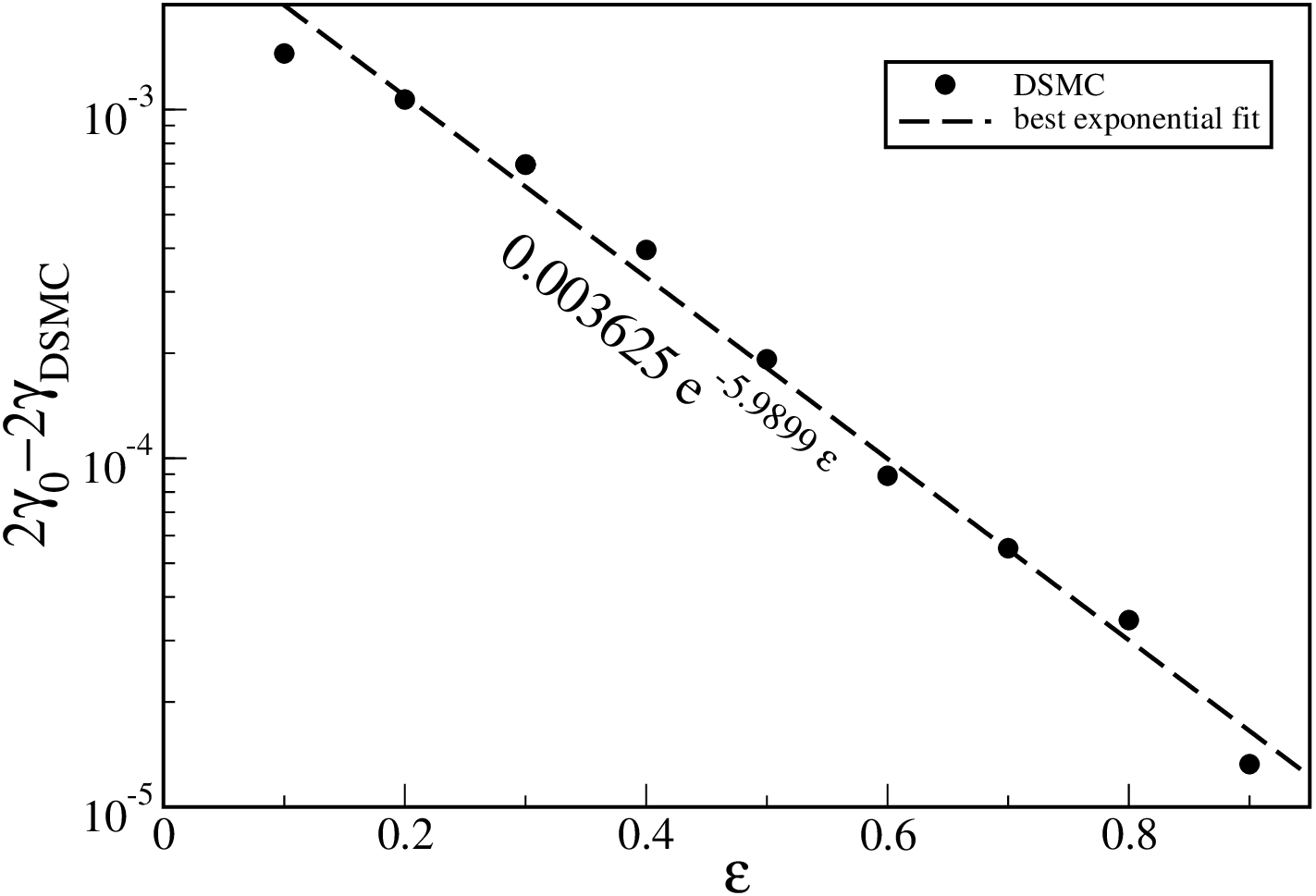}}
\caption{Difference of the cooling coefficient due to
Eq. (\ref{eq:gamma_0}) and the numerically values,
$(\gamma_0-\gamma_{\rm DSMC})$, over the coefficient of
restitution. Since Eq. (\ref{eq:gamma_0}) disregards the exponential
tail of the distribution, the shown difference is due to the presence
of the tail. The dashed line shows the best exponential fit.}
\label{fig:gamma}
\end{figure}

The dependence of $\gamma_0$ on the coefficient of restitution
$\varepsilon$ is shown in Fig. \ref{fig:gamma0} together with the
numerical result.
The numerical results (see below) and the theoretical results
disregarding the tail, Eq. (\ref{eq:gamma_0}), virtually
coincide. When we plot the difference of both curves, however, we find a
systematic deviation between the numerical values and the analytical
result (Fig. \ref{fig:gamma}).

\begin{figure}[t]
\centerline{\includegraphics[width=8cm,clip]{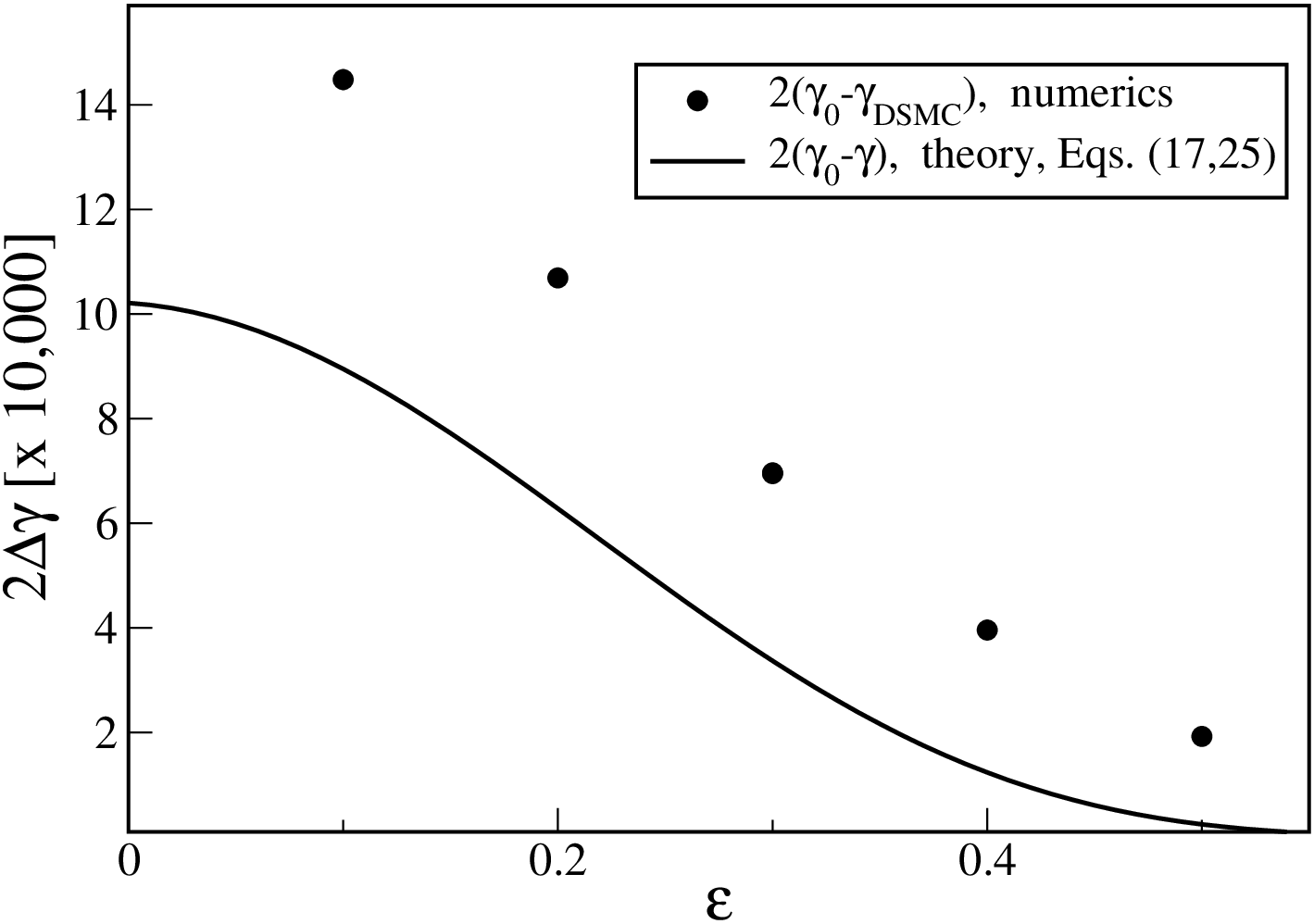}}
\caption{Influence of the tail on the cooling coefficient. The symbols
show the same data as in Fig. \ref{fig:gamma} (but in linear
scale). The full line shows $2(\gamma_0-\gamma)$ as given by
Eqs. (\ref{eq:gamma_0}) and (\ref{eq:gamma}).}
\label{fig:gammaDiffTheory}
\end{figure}

To estimate this rather small difference quantitatively, we write the
distribution function as a sum of two terms,
\begin{equation}
\label{eq:f_ft_f0}
\tilde{f}(c)= \tilde{f}_0(c)+ \tilde{f}_{\rm t} (c) \, , 
\end{equation}
that is, the main part $\tilde{f}_0(c)$ (with $c$ extended to
infinity), and the pure overpopulation of the tail, $\tilde{f}_t (c)$,
\begin{eqnarray}
\label{eq:f_with tail}
&&\tilde{f}_0(c)= A c^2 e^{-c^2} \left[ 1 +a_2 S_2\left(c^2\right) \right] \\
&& \tilde{f}_t (c)= \left[ B c^2 \exp (-b\, c ) - \tilde{f}_0(c) \right] \Theta \left(c - c_*\right) \,.
\end{eqnarray}
The product $\tilde{f}(c_1) \tilde{f}(c_2) $ in Eq. (\ref{eq:kin_int}) reads then
\begin{equation}
\label{eq:fc1fc2}
 \tilde{f}(c_1) \tilde{f}(c_2) = \tilde{f}_0(c_1)\tilde{f}_0(c_2) +
 \tilde{f}_0(c_1) \tilde{f}_{\rm t}(c_2) + 
\tilde{f}_{\rm t}(c_1) \tilde{f}_0(c_2) + 
\tilde{f}_{\rm t}(c_1) \tilde{f}_{\rm t}(c_2) \nonumber \,.
\end{equation}
For large $c_*$ one can neglect the last term in the above equation. Moreover, in this case
we approximate $\tilde{f}_{\rm t}(c)\approx B\exp(-b\, c)$ and
$\vec{c}_{12} \cdot \vec{e}\approx \vec{c}_{1} \cdot \vec{e}$,
assuming $c_1 \gg c_2$. A similar approximation may be applied for the
opposite case, $c_2 \gg c_1$.  Taking then into account the symmetry
of the integrand in Eq. (\ref{eq:kin_int}) with respect to $\vec{c}_1$
and $\vec{c}_2$, we finally obtain an estimate for $J_k$:
\begin{equation}
\label{eq:approx_J}
\begin{aligned}
J_k &= \frac{\pi}{16}A^2 J_k^{(0)}   
  +2 \left[ \int d \vec{c}_2 \tilde{f}_0 (c_2) \right] \int d
  \vec{c}_1 \int d \vec{e}\,\, \Theta(-\vec{c}_{1} \cdot \vec{e}\, )
  \left|\vec{c}_{1} \cdot \vec{e}\,\right|^{k+1}  \tilde{f}_{\rm t}(c_1)  \\ 
& = \frac{\pi}{16}A^2 J_k^{(0)}   
 +2A \frac{\sqrt{\pi}}{4}  \, 4 \pi  \int_{c_*}^{\infty} c_1^{k+3}  B e^{-bc_1} dc_1 2 \pi 
\int_{\frac{\pi}{2}}^{\pi} \sin \theta \left| \cos \theta \right|^{k+1}  \, , 
\end{aligned}
\end{equation}
where $J_k^{(0)}$ is the value of the coefficient $J_k$ for the
distribution function $\tilde{f}_0(c)$ with $A$ in Eq. (\ref{eq:f_with tail}) equal to $4/\sqrt{\pi}$, that is,
for the case when the exponential tail is disregarded. In particular, we need
\begin{equation}
\label{eq:Jk0}
\begin{aligned}
J_0^{(0)} &= 2 \sqrt{2 \pi} \left(1 - \frac{1}{16}a_2 \right) \\ 
J_2^{(0)} &= 4 \sqrt{2 \pi} \left(1 +  \frac{3}{16}a_2  \right) \, . 
\end{aligned}
\end{equation}
Performing the integration in Eq. (\ref{eq:approx_J}) we obtain 
\begin{equation}
\label{eq:Jk_final}
J_k = \frac{\pi}{16}A^2 J_k^{(0)}  
  + \frac{\sqrt{\pi}}{4} \frac{2 AB}{(k+2)} \frac{8 \pi^2}{b^{k+4}} \Gamma(k+4, b c_*)  \, , 
\end{equation}
where $\Gamma(x,y)$ is the incomplete gamma-function. Inserting into
Eq. (\ref{eq:gam_gen}) yields the cooling rate,
\begin{equation}
\label{eq:gamma}
\gamma = \frac{(1-\varepsilon^2)}{6} \, 
 \frac{\left(1+\frac{3}{16}a_2 \right)Ab^6  + 2 \pi \sqrt{2} B \Gamma(6, bc_*)}
{\left(1-\frac{1}{16}a_2 \right)Ab^6 + 8 \pi \sqrt{2} B b^2 \Gamma(4, bc_*)}  \, . 
\end{equation}
Naturally, for $B=0$, corresponding to absence of the tail (see
Eq. (\ref{eq:ftentat})), Eq. (\ref{eq:gamma}) reduces to the previous
relation Eq. (\ref{eq:gamma_0}). In Fig. \ref{fig:gammaDiffTheory} the
difference $(\gamma_0 - \gamma)$ which quantifies the contribution to
the cooling coefficient from the tail is compared with the numerical
results.

The scaling law, $\gamma_0-\gamma_{\rm DSMC}\propto
\exp(-6\varepsilon)$, shown in Fig. \ref{fig:gamma} as a numerical
result is, however, difficult to confirm analytically. The approximate
analytical theory shown in Fig. \ref{fig:gammaDiffTheory} agrees with
the numerical results only qualitatively.

\section{Slow relaxation of the velocity distribution}

In the above analysis we assumed that the velocity distribution
 function $\tilde{f}(c)$ had already relaxed to its steady state
 scaling shape. In our numerical experiments we observed, however,
 that the relaxation to the stationary form occurs extremely slowly,
 as compared to the relaxation of molecular gases to equilibrium
 distribution. To study this retarded relaxation quantitatively, we
 use the coefficient $\gamma_{\rm DSMC}$ described above, and define
 the temperature $T_{\rm fit}(\tau) \propto \exp(-2\gamma_{\rm DSMC}
 \, \tau)$. By definition, for $\tau \gg 1$ one obtains $T_{\rm
 DSMC}\approx T_{\rm fit}$ since $\gamma_{\rm DSMC}$ was determined as
 the best exponential fit to $T_{\rm DSMC}(\tau)$ for $\tau \gg
 1$. Hence, the quantity $1-T_{\rm DSMC}(\tau)/T_{\rm fit}(\tau)$
 characterizes the relaxation of the distribution function to its
 stationary form. We observe that the relaxation time depends
 sensitively on the coefficient of restitution $\varepsilon$.

\begin{figure}[b]
\centerline{\includegraphics[width=8cm,clip]{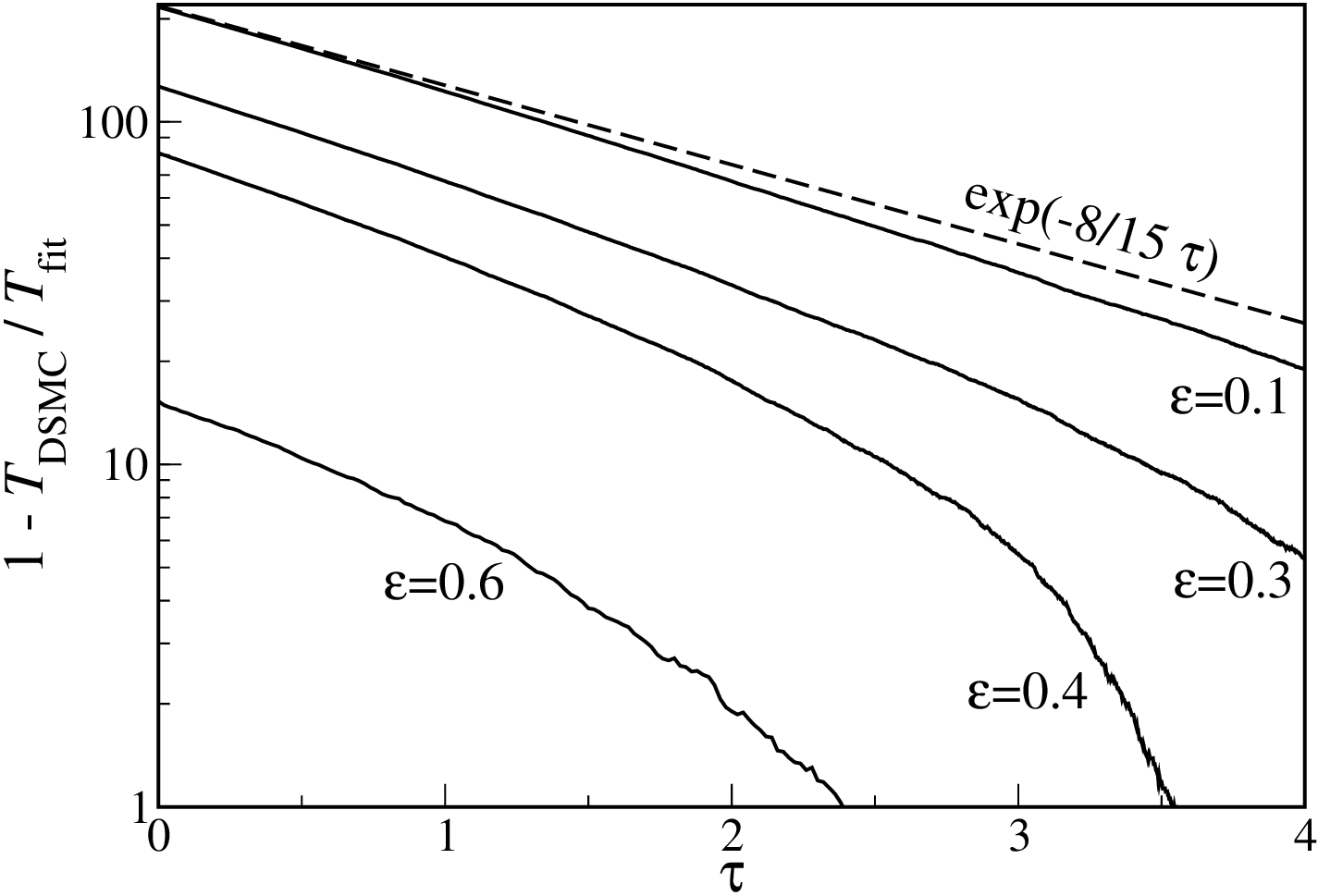}}
\caption{Short-time relaxation of the temperature decay to Haff's law,
Eq. (\ref{eq:dTdt}) for a system of $N=10^8$ particles for different
coefficients of restitution, $\varepsilon=0.1$, $\varepsilon=0.3$,
$\varepsilon=0.4$, and $\varepsilon=0.6$. The initial relaxation
occurs on the time scale $1-T_{\rm DSMC}/T_{\rm fit}\propto \exp(-8/15
\tau)$.}
\label{fig:Trelax1}
\end{figure}

\begin{figure}[b]
\centerline{\includegraphics[width=8cm,clip]{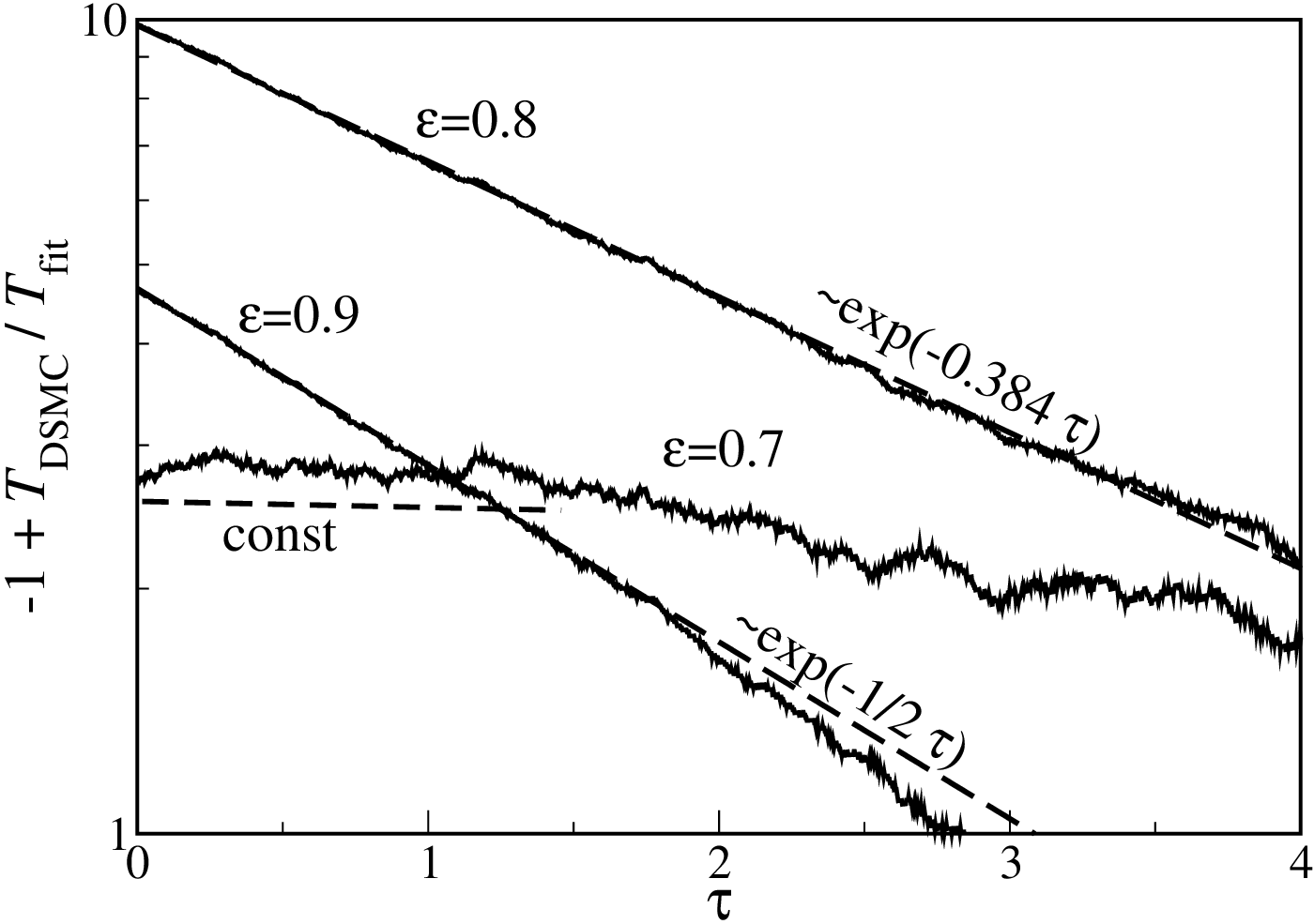}}
\caption{Same as Fig. \ref{fig:Trelax1} but for $\varepsilon=0.7$,
$\varepsilon=0.8$, and $\varepsilon=0.9$. Note that the vertical axis
has opposite sign as the axis in Fig. \ref{fig:Trelax1}.}
\label{fig:Trelax2}
\end{figure}

\begin{figure}[b]
\centerline{\includegraphics[width=8cm,clip]{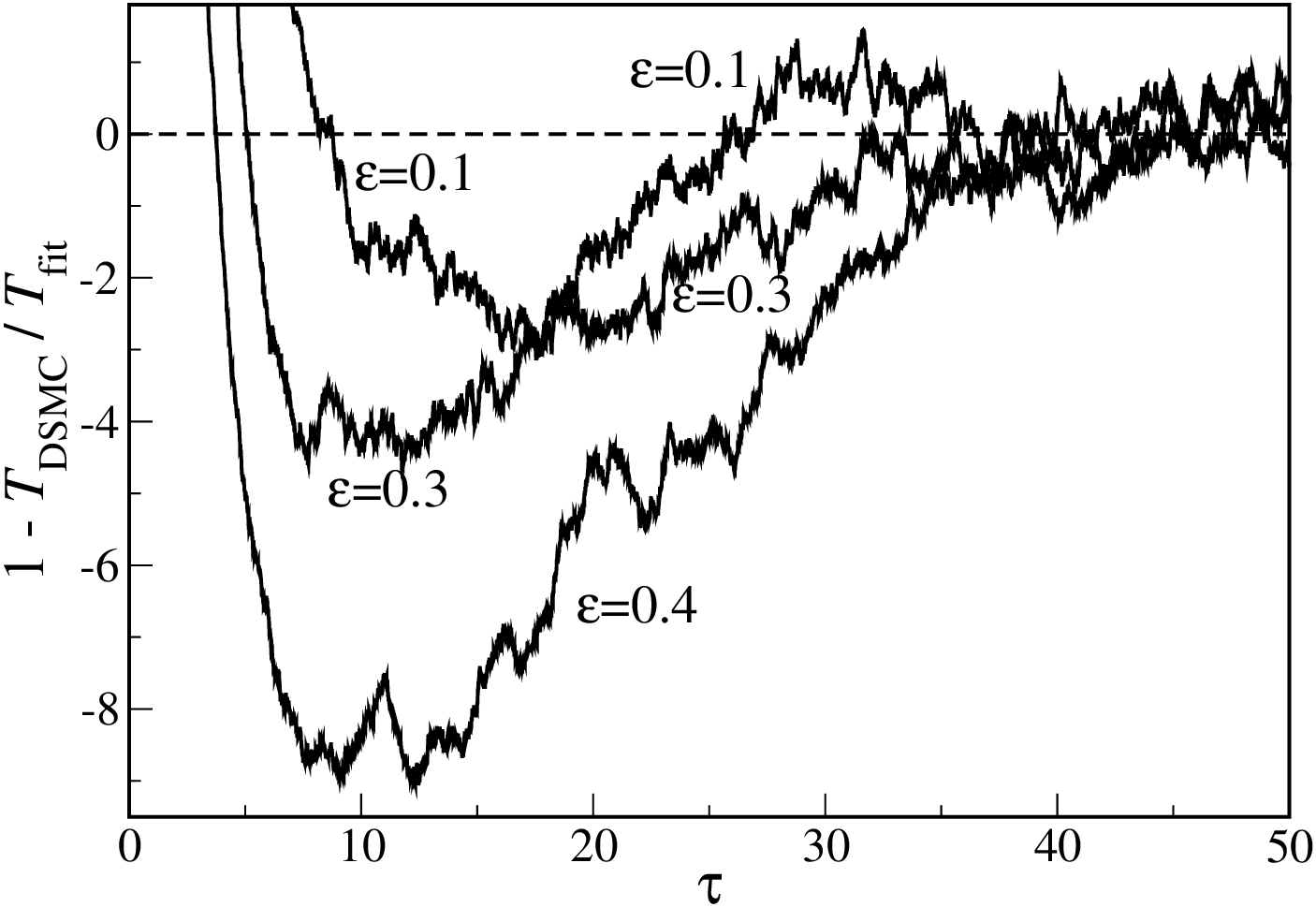}}
\caption{Long-time relaxation of the temperature decay rate for $\varepsilon=0.1$, $\varepsilon=0.3$, and $\varepsilon=0.4$.}
\label{fig:Trelax3}
\end{figure}

\begin{figure}[b]
\centerline{\includegraphics[width=8cm,clip]{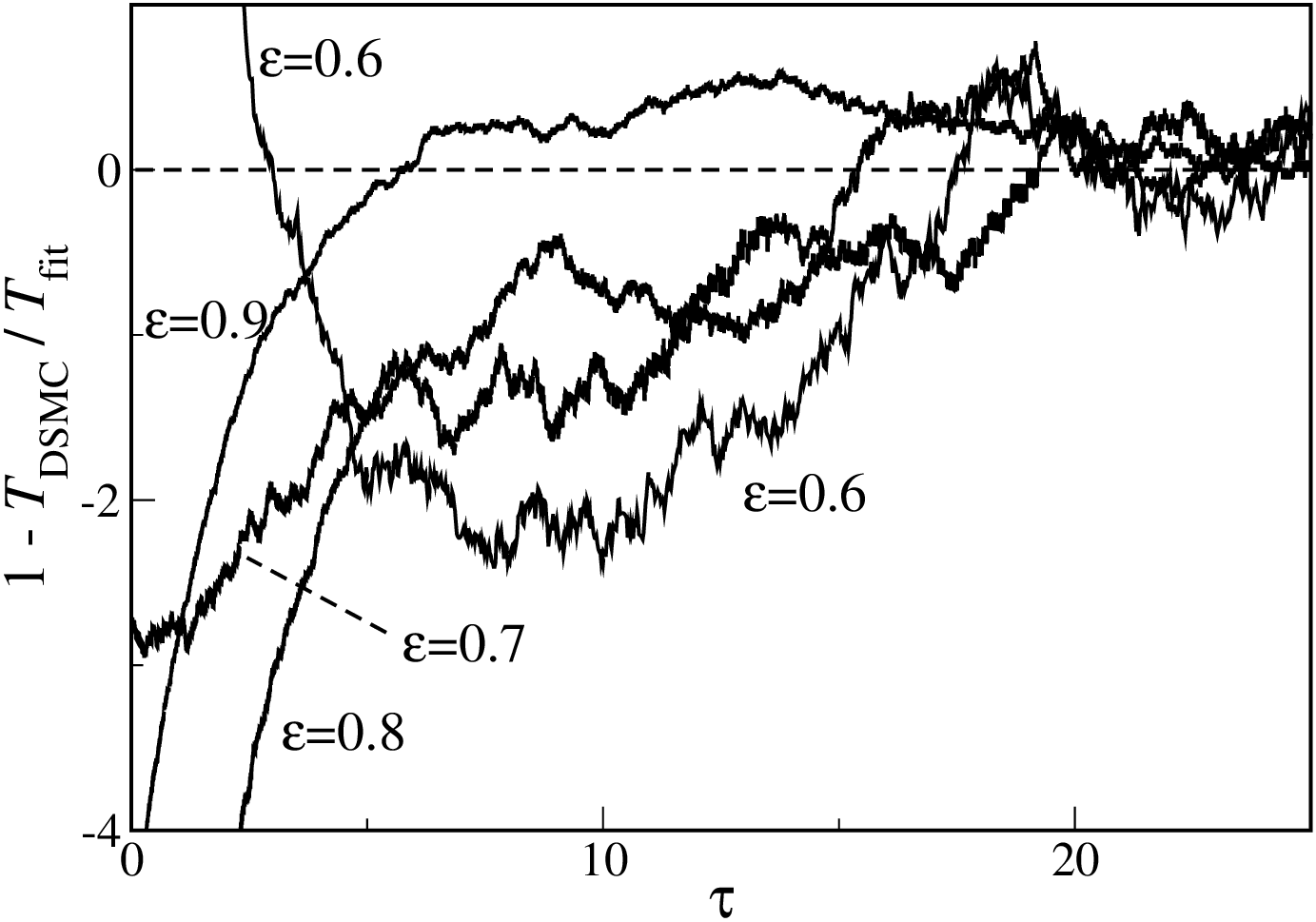}}
\caption{Same as Fig. \ref{fig:Trelax3} but for $\varepsilon=0.6$, $\varepsilon=0.7$, $\varepsilon=0.8$, and $\varepsilon=0.9$.}
\label{fig:Trelax4}
\end{figure}

Figures \ref{fig:Trelax1}-\ref{fig:Trelax4} illustrate the relaxation
kinetics for different values of the coefficient of restitution
$\varepsilon$. In Figs. \ref{fig:Trelax1} and \ref{fig:Trelax2} the
initial relaxation is shown.  We observe a initial quick relaxation
with characteristic time of 3-5 collisions per particles, similar as
for molecular gases.

For values $\varepsilon <0.7$ the simulated temperature
approaches Haff's law from below (Fig. \ref{fig:Trelax1}), whereas for
$\varepsilon>0.7$ it approaches Haff's law from above (note the
different labeling of the vertical axis in these figures). For
$\varepsilon =0.7$ the initial relaxation rate vanishes. This may be
explained by the fact that at $\varepsilon\approx 0.7$ the value of
the second Sonine coefficient changes its sign, see
Eq. (\ref{eq:a2NE}). That is, for $\varepsilon\lesssim 0.7$ the
stationary velocity distribution $\tilde{f}$ is bent towards lower
velocities as compared with the Maxwell distribution while for
$\varepsilon\gtrsim 0.7$ the distribution is bent towards higher
velocities. For $\varepsilon =0.7$ the second Sonine coefficient is
very small, $a_2(0.7)\approx0$, therefore, for $c\sim 1$ the
distribution function is very close to the Maxwell distribution and,
thus, there is no initial relaxation. These arguments prove that the
initial relaxation shown in Figs. \ref{fig:Trelax1} and
\ref{fig:Trelax2} corresponds to the relaxation of the main part of
the velocity distribution, $c\sim 1$, whose deviation from the Maxwell
distribution is described by the second Sonine coefficient $a_2$.

For small coefficient of restitution,  $\varepsilon \le 0.6$, the initial slope of the relaxation curves is almost independent of $\varepsilon$. Its value is surprisingly close to the slope $-8/15$ of the relaxation curve of the second Sonine coefficient in the limit of {\em almost elastic} particles, $\varepsilon\lesssim 1$.\cite{BrilliantovPoeschelOUP} Also surprisingly, for larger values of  the restitution coefficients $\varepsilon >0.7$, which correspond to negative values of the second Sonine coefficient, the slope becomes smaller and depends noticeably on  $\varepsilon$. For almost elastic particles, $\varepsilon=0.9$ we find the initial slope 1/2. Presently we do not have an explanation for this behavior of the initial relaxation.  

In Figs. \ref{fig:Trelax3} and \ref{fig:Trelax4} we present the
complete relaxation to the steady state.
The complete relaxation requires much longer time as compared to
molecular gases. This is related to the relatively slow formation of
the exponential tail. The relaxation time depends on the
coefficient of restitution as shown in Fig. \ref{fig:Trelax3}. The
larger the coefficient of restitution the longer it takes to form the
complete velocity distribution including the exponential tail.

To explain this effect we use the equation for the relaxation of the
velocity distribution function to its scaling
form,\cite{BrilliantovPoeschelOUP,BrilliantovPoeschel:2000visc}
\begin{equation}
\label{eq:kin_eq_for_fc}
\frac{\mu_2}{3} \left( 3 + c \frac{\partial}{\partial c } 
\right)  \tilde{f}(c,\tau )+ J_0 \frac{\partial}{\partial \tau  } \tilde{f}(c,\tau) = \tilde{I}
\left( \tilde{f},  \tilde{f} \right)
\end{equation}
where $\tilde{I} ( \tilde{f}, \tilde{f} )$ is the reduced collision
integral, Eq. (\ref{eq:I_scale}), and $J_0$ is given in
Eq. (\ref{eq:kin_int}). As usual we neglect the incoming term for $c
\gg 1$\cite{EsipovPoeschel:1995,NoijeErnst:1998}, leading to the
approximation
\begin{equation}
\label{eq:coll_int}
\tilde{I} \left( \tilde{f} , \tilde{f} \right) \approx - \pi c \tilde{f}(c) ~~~~~\mbox{for}~~~~~
c \gg 1 \, . 
\end{equation}
With the Ansatz $\tilde{f}(c,\tau ) =B \exp \left[ -w(\tau ) c\right]$
we recast Eq. (\ref{eq:kin_eq_for_fc}) into
\begin{equation}
\label{eq:dwdt}
\frac{d w}{d \tau} + \frac{\mu_2}{3J_0}  w = \frac{\pi}{J_0}  ~~~~~\mbox{for}~~~~~~
c \gg 1 \, 
\end{equation}
with the solution
\begin{equation}
\label{eq:wt_result}
w( \tau ) = b + (1-b) \exp\left[-\frac{\tau}{\tau_0(\varepsilon)} \right]  \, ,
\end{equation}
where $b= 3 \pi / \mu_2$ coincides with Eq. (\ref{eq:tail}) and
$\tau_0^{-1}(\varepsilon) = \mu_2/3 J_0$. Neglecting $a_2$, which
quantifies (small) deviations of the main part of the distribution
with respect to the Maxwellian, and the contribution from the tail,
Eq. (\ref{eq:Jk0}) yields $J_0 =2 \sqrt{2 \pi}$, which together with  Eq. (\ref{eq:mu2}) leads to the relaxation time
\begin{equation}
\tau_0^{-1}(\varepsilon) = \frac{1-\varepsilon^2}{6} \, .
\label{eq:tau0}
\end{equation}
For example, for $\varepsilon =0.4$ we obtain the theoretical
relaxation time, $\tau_0 \approx 7.1$. As shown in Fig.
\ref{fig:Trelax3} for $\varepsilon=0.4$ in the simulation the quantity
$(1-T_{\rm DSMC}/T_{\rm fit})$ decreases by the factor of $10$ in the
time span $\Delta \tau = 25$, ranging from $\tau=10$ to
$\tau=35$. This gives the estimate for the relaxation time
$\tau_0=25/\log(10) \approx 10.8$, in agreement with the above
theoretical value $\tau_0=7.1$.

From the above theory we expect that the relaxation time increases
with $\varepsilon $.  While this tendency is confirmed for small
coefficients of restitution from $\varepsilon =0.1$ to $\varepsilon
=0.4$, Fig. \ref{fig:Trelax3}, the relaxation time seems to saturate
for larger $\varepsilon \ge 0.6$, Fig. \ref{fig:Trelax4}. This is,
presumably, a finite size effect: For large values of $\varepsilon$
the number of particles in the tail, which is determined by the
threshold velocity $c_*$, increasing with $\varepsilon$, is not large
enough to develop a significant tail. Therefore for such systems the
apparent relaxation occurs much faster then it would be in a
sufficiently large system. The relaxation time is mainly determined by
the number of particles in the tail, rather then by the coefficients
of restitution $\varepsilon$. Nevertheless, even for these systems,
which are not sufficiently large for a quantitative study of the tail
relaxation, the latter occurs anomalously slow.

\section{Conclusion}

We studied analytically and numerically the impact of the high-energy
tails of the velocity distribution function in granular gases in the
homogeneous cooling state on the cooling coefficient and on the
relaxation time towards the steady state velocity distribution.
 
In our analytical theory we used an universal Ansatz for the velocity
distribution function for the entire range of velocities. This Ansatz
comprises the main part of the distribution function, where the
velocities are of the order of the thermal velocity, $v \sim v_T$ as
well as the tail, where $v \gg v_T$.

We derived the coefficients of the proposed Ansatz, which allows to
estimate the cooling coefficient and to characterize the impact of the
high-energy tail on the cooling. Our analytical results are in
qualitative agreement with numerical data obtained by Direct
Simulation Monte Carlo of $10^8$ particles.

In the simulations we found anomalously slow relaxation of the
velocity distribution function to its stationary form as compared to
the relaxation of molecular gases, where the relaxation occurs during
few collision times. Instead, for granular gases, we observed much
longer relaxation, in the order of 20-30 collisions per
particle. To explain this slow relaxation we developed a theory, which
predicts that the relaxation time increases with decreasing
inelasticity, in agreement with the numerical observations. For a large
coefficient of restitution the relaxation time saturates with
increasing $\varepsilon$, which may be attributed to finite size
effects, that is, the system of $10^8$ particles is not large
enough for the quantitative numerical analysis of the tail relaxation.


\begin{thebibliography}{10}
\expandafter\ifx\csname url\endcsname\relax
  \def\url#1{\texttt{#1}}\fi
\expandafter\ifx\csname urlprefix\endcsname\relax\def\urlprefix{URL }\fi

\bibitem{BritoErnst:1998}
R.~Brito, M.~H. Ernst, Extension of {H}aff's cooling law in granular flows,
  Europhys. Lett. 43 (1998) 497.

\bibitem{GoldhirschZanetti:1993}
I.~Goldhirsch, G.~Zanetti, Clustering instability in dissipative gases, Phys.
  Rev. Lett. 70 (1993) 1619.

\bibitem{MontaneroSantosTailsGG:1999}
J.~M. Montanero, A.~Santos, Computer simulation of uniformly heated granular
  fluids, Granular Matter 2 (1999) 53--64.

\bibitem{EsipovPoeschel:1995}
S.~E. Esipov, T.~P\"oschel, The granular phase diagram, J. Stat. Phys. 86
  (1997) 1385.

\bibitem{Haff:83}
P.~K. Haff, Grain flow as a fluid-mechanical phenomenon, J. Fluid Mech. 134
  (1983) 401.

\bibitem{GoldshteinShapiro1:1995}
A.~Goldshtein, M.~Shapiro, Mechanics of collisional motion of granular
  materials. {P}art 1: {G}eneral hydrodynamic equations, J. Fluid Mech. 282
  (1995) 75.

\bibitem{NoijeErnst:1998}
T.~P.~C. van Noije, M.~H. Ernst, Velocity distributions in homogeneous granular
  fluids: the free and the heated case, Granular Matter 1 (1998) 57.

\bibitem{GoldhirschetalNLP:2003}
I.~Goldhirsch, H.~S. Noskowicz, O.~Bar-Lev, The homogeneous cooling state
  revisited, in: T.~P\"oschel, N.~V. Brilliantov (Eds.), Granular Gas Dynamics,
  Vol. 624 of Lecture Notes in Physics, Springer, Berlin, 2003, pp. 37 -- 63.

\bibitem{PoeschelBrilliantovFormella:2006}
T.~P\"oschel, N.~V. Brilliantov, A.~Formella, Impact of high-energy tails on
  granular gas properties, Phys. Rev. E 74 (2006) 041302.

\bibitem{BrilliantovPoeschelOUP}
N.~V. Brilliantov, T.~P\"oschel, Kinetic Theory of Granular Gases, Oxford
  University Press, Oxford, 2004.

\bibitem{Bird:1994}
G.~A. Bird, Molecular Gas Dynamics and the Direct Simulation of Gas Flows,
  Oxford University Press, 1994.

\bibitem{MontaneroSantos:1996}
J.~M. Montanero, A.~Santos, Monte {C}arlo simulation method for the {E}nskog
  equation, Phys. Rev. E 54 (1996) 438.

\bibitem{BrilliantovPoeschel:2000visc}
N.~V. Brilliantov, T.~P\"oschel, Velocity distribution of granular gases of
  viscoelastic particles, Phys. Rev. E 61 (2000) 5573.

\end{thebibliography}

\end{document}